\documentclass[twocolumn,showpacs,preprintnumbers,amsmath,amssymb]{revtex4}

\usepackage{graphicx}
\usepackage{dcolumn}
\usepackage{bm}

\begin{document}

\preprint{APS/123-QED}

\title{Rejection-free Monte Carlo Algorithms for Models with
  Continuous Degrees of Freedom}
\author{\small J.D. Mu\~noz$^{1}$, M.A. Novotny$^{2}$
  and S.J. Mitchell$^{2,3}$\\
\small${}^{1}$Institute for Computer Applications 1, University of Stuttgart,\\
\small Pfaffenwaldring 27, D-70569 Stuttgart, Germany\\
\small Permanent address:
\small Dpto. de F\'{\i}sica, Univ. Nacional de Colombia,\\
\small Bogota D.C., Colombia\\
\small${}^{2}$ Computational Science and Information Technology, 
Florida State University,\\
\small Tallahassee, Florida 32306-4120, USA\\
\small${}^{3}$ Center for Materials Research and Technology and 
 Department of Physics,\\
 Florida State University, Tallahassee, Florida 32306-4351, USA
}

\date{\today}

\begin{abstract}
We construct a rejection-free Monte Carlo algorithm for a system with continuous degrees of freedom.  
We illustrate the algorithm by applying it to the classical three-dimensional Heisenberg model 
with canonical Metropolis dynamics. 
We obtain the lifetime of the metastable state following a reversal of the external magnetic field. 
Our rejection-free algorithm obtains results in agreement with a direct implementation of the Metropolis dynamic and requires orders
of magnitude less computational time at low temperatures.  
The treatment is general and can be extended to other dynamics and other systems with continuous degrees of freedom.
\end{abstract}


\maketitle


Nucleation and metastability are characteristic behaviors of dynamical processes for many different fields, 
from stock markets and sociology \cite{StaufferEvol} 
to parallelization methods for massively parallel computers \cite{Novotny2000}
to chemical reactions and materials science.
Many classical models, 
when simulated with Monte Carlo methods, 
also present these behaviors, 
and although the number of steps in the simulation does not necessarily correspond directly to experimental time, 
they give valuable insights into these dynamic phenomena. 
For instance, 
studies on Ising \cite{DropsIsing} and anisotropic Heisenberg models \cite{DropsHeisenberg} 
have shown the existence of different metastable decay regimes for small ferromagnetic particles 
after a reversal of the external magnetic field. 
At low temperatures,
single and multiple droplet nucleation and a strong field regime are observed,
and recently, 
indirect experimental evidence of these regimes has been found \cite{DropsExp}.

Many Monte Carlo dynamics are Markov processes that divide each step into two successive parts: 
first, a new state is chosen,
second, it is accepted or rejected according to some criteria.
In many cases of interest, 
the acceptance rates can be so small that a huge number of trials is required before a change is made. 
Then, a direct implementation of the Monte Carlo dynamic,
one that attempts steps one after the other, 
is extremely slow.
For instance, 
Ising models with Metropolis dynamics \cite{Metropolis} can require $10^{15}$ trials to leave a metastable state at low temperatures, 
and such a simulation would take $10^{10}$ minutes \cite{Novotny1}.
Therefore, techniques to implement the same dynamic in a faster way are required. 

There exist different techniques to construct rejection-free implementations of Monte Carlo dynamics for discrete spin systems,
like the $n$-fold way \cite{Lebowitz,Novotny2} and Monte Carlo with absorbing Markov chains \cite{Novotny1} algorithms 
(for a review see \cite{DynAlgorithms}). 
In this paper, 
we extend the $n$-fold way rejection-free technique to systems with continuous degrees of freedom,
and we construct a rejection-free algorithm for the classical Heisenberg model.
Our treatment is completely general and can be extended to other dynamics and other continuous systems.

%
\label{ImportanceSamping}

Consider a Markov process with every step consisting of two parts.
First, choose a movement from state $x$ to state $x' \neq x$ with probability $T(x'|x)$.
Second, accept it with probability $A(x'|x)$. 
The full probability to undergo the movement $(x'|x)$ is then
\begin{equation}
  \label{WTA}
  W(x'|x) = T(x'|x)  A(x'|x)  \; .
\end{equation}
In this paper,
the term ``direct implementation'' refers to the common selection and rejection implementation of the Markov process,
in which two random numbers are used, 
one for the selection of $x'$ and one for the rejection/acceptance of $x'$.

In the specific case of importance sampling,
$T(x'|x)$ is chosen to be symmetric, 
$T(x'|x)=T(x|x')$, 
and $A(x'|x)$ is tuned to obtain a desired limit probability distribution, $P(x)$, 
when the number of steps tends to infinity.
This is accomplished by requiring the detailed balance condition,
\begin{equation}
  \label{DetailedBalance}
  W(x'|x) P(x) = W(x|x') P(x')  \; .
\end{equation}
To obtain the canonical distribution, $P(x) \propto \exp[-E_x/k_BT]$,
a widely used choice for $A(x'|x)$ is the Metropolis acceptance probability
\begin{equation}
  \label{Metropolis}
    A(x'|x)=\nu \min{\left\{ 1, \exp[-(E_{x'}-E_x)/k_BT] \right\}} \; ,
\end{equation}
where $\nu$ is a constant which is the same for all $x$ and $x'$ 
and is included to ensure that $\sum_{x'} W(x'|x)\le 1$ for all algorithmic steps.
For all discussions presented here, $\nu=1$ was sufficient to ensure this condition.
The usual Metropolis simulation \cite{Metropolis} is a direct implementation of this dynamic.
 
\label{N_Fold-Way}

The $n$-fold way \cite{Lebowitz,Novotny2} is a rejection-free implementation of the dynamic $W(x'|x)$,
and we briefly remind the reader of this two-part algorithm.
First, the number of trials, $t$, to leave the current state is computed (the update time),
and second, one movement is chosen and performed.
In this way, every algorithmic step induces a change in the system configuration, 
but the dynamic information is preserved.

The first step of the $n$-fold way algorithm consists of generating the update time, $t$,
which is a random variable chosen from the appropriate probability distribution.
Define $\lambda$ as the probability to reject all movements,
\begin{equation}
  \label{lambda}
  \lambda = 1-\sum_{x'} W(x'|x) \; ,
\end{equation}
and the probability $p(t)$ to leave state $x$ after $t$ steps is a geometric distribution \cite{Novotny2},
\begin{equation}
  \label{pt}
  p(t)= \lambda^{t-1}(1-\lambda) \; .
\end{equation}
A so-called integral generator \cite{NumericalRecipies} can be constructed to produce $t$ with this distribution. 
Define $I(t)$$:=$$1-\sum_{k=1}^t p(k)$$=$$\lambda^t$ ($I(0)$$:=$$1$), 
and let $\tilde{r}$ be a random number uniform on $(0,1)$.
The number of steps until the next update, $t$, is then determined by
\begin{equation}
  \label{t}
  I_{t-1} \leq \tilde{r} < I_t \; \mbox{\rm and } \; 
  t=\left\lfloor  {\ln \tilde{r} \over \ln \lambda} \right\rfloor +1 \; ,
\end{equation}
where $\lfloor x \rfloor$ denotes the integer part of $x$.

In the second step of the $n$-fold way algorithm, 
we must choose the exit state, $x'$, from the appropriate probability distribution.
The probability to exit to state $x'$ from $x$ is
\begin{equation}
  \label{C}
      C(x'|x)={1 \over 1-\lambda} W(x'|x) \; .
\end{equation}
A movement is chosen with this probability by using the same integral-generator strategy as in the first step of the algorithm.
The movements are ordered with index $k$ (discrete degrees of freedom),
and the partial sums $Q(k)$$:=$$\sum_{m=1}^k C(x'|x)_m$, with $Q(0)$$:=$$0$, are computed.
The movement $k$ is chosen if $Q(k-1)$$\leq$$\tilde{r}$$<$$Q(k)$,
where $\tilde{r}$ is uniformly distributed on $(0,1)$.
To perform the movement selection in a more efficient way, 
movements are grouped into {\it classes}, 
and a two-level search is performed by first selecting the class $i$ with probability $C(i|x)$$=$$\sum_{x' \in i}C(x'|x)$ 
and then selecting the movement $(x'|i)$ from inside the class with probability $C(x'|i)$$=$$C(x'|x)/C(i|x)$ (Bayes). 
This completes the $n$-fold algorithm,
which is a rejection-free Monte Carlo algorithm for systems with discrete degrees of freedom.

In importance sampling for discrete spin systems \cite{Lebowitz,Novotny2},
the exit states, $x'$, are usually grouped into classes by energy changes,
where all movements in class $i$ have the same energy change, $\Delta E$, and
thus the same acceptance rate, $A_i$. 
Let $n_i$ be the number of movements in class $i$.
Using $T(x'|x)$$=$$1/\hat{N}$, where $\hat{N}$ is the total number of possible movements,
we obtain the classic $n$-fold way expressions \cite{Lebowitz}
{\small \begin{equation}
  \lambda= 1-\sum_i {n_i \over \hat{N}} A_i \; ; \; 
  C(i|x)={n_i A_i \over (1-\lambda)\hat{N}} \; ; \; C(x'|i)={1 \over n_i} \; .
\end{equation}}

To apply this rejection-free technique to systems with continuous degrees of freedom, 
some ideas employed to extend the broad histogram method for continuous systems are used \cite{BHM}.
Discrete probability distributions become probability density functions (pdf),
and a discrete choice of a random variable becomes the
construction of a random generator with its proper distribution.
To illustrate this process, 
we construct a rejection-free algorithm for the classical Heisenberg model with Metropolis dynamics. 
The Hamiltonian is
\begin{equation}
{\small 
  \label{Heisenberg}
    {\mathcal H}= -\sum_{\langle {\it i j} \rangle} 
    \left\{J_xX_iX_j + J_yY_iY_j + J_zZ_iZ_j\right\} -H_z \sum_i Z_i \; , 
}
\end{equation}
where $\vec{\sigma}_i$$=$$X_i \hat{x} + Y_i \hat{y} + Z_i \hat{z}$ is a spin of unit length on site $i$, 
$H_z$ is the magnitude of an external magnetic field in the $z$-direction, 
$\langle ij \rangle$ represents a nearest-neighbor summation, 
and $J_x$, $J_y$, and $J_z$ are coupling constants. 

For continuous systems like the Heisenberg model, 
the movements form an uncountable set. 
With $x$ fixed, $T(x'|x)$ can be interpreted as a pdf in the configuration space of the system, 
where $T(x'|x) dx'$ is the probability to choose the new state, $x'$, inside a small
infinitesimal region $dx'$, centered on $x'$.
Let us fix $T(x'|x)$ by choosing all movements with the same probability as follows. 
First, a site $i$ is chosen with probability $1/N$,
where $N$ is the number of sites.
Second, a new orientation $\vec{\sigma}'_i$ for the spin at $i$
is chosen uniformly on the unit sphere
(pdf $T(\theta',\varphi'|i)$$=$$1/4 \pi$).
This is equivalent to generating $z'$$\equiv$$\cos \theta'$ uniformly on the
interval $[-1,1]$ and $\varphi'$ uniformly on the interval $[0,2\pi)$,
where $(z',\varphi')$ are the coordinates of $\vec{\sigma}'_i$ in some cylindrical coordinate system \cite{BHM}.
Let $(z,\varphi)$ be the coordinates of $\vec{\sigma}_i$ in the same system .
The total pdf of this movement $(z',\varphi'|z,\varphi)_i$ is,
therefore,  $T(z',\varphi'|z,\varphi)_i$$=$$1/4 \pi N$,
and clearly, $T(x|x')$$=$$T(x'|x)$.

The energy change of this movement is 
\begin{equation}
{\small 
  \label{DE}
  \Delta E = (\vec{\sigma}_i - \vec{\sigma}'_i) \cdot \vec{S_i}
  \; ,
}
\end{equation}
\begin{equation*}
{\small 
  \vec{S_i} = \left[ J_x \sum_{j} X_j \right] \hat{x} 
  + \left[ J_y \sum_{j} Y_j \right] \hat{y} 
  + \left[ H_z + J_z \sum_{j} Z_j \right] \hat{z} \; .
}
\end{equation*}
Here, $j$ denotes a sum over the nearest neighbors of site $i$.
If we rotate to a coordinate system with the $z$-axis parallel to $\vec{S}_i$, 
this energy change reduces to $\Delta E$$=$$-(z'-z)S_i $. 
Therefore, from Eq.~(3),
\begin{equation}
{\small 
  A(z',\varphi'|z,\varphi)_i = \left\{ \begin{array}{lcl}
      \exp[(z'-z)S_i/k_BT] &;& \mbox{\rm if } z'<z \\
      1 &;& \mbox{\rm otherwise}
      \end{array} \right. \; .
}
\end{equation}
Thus, $W(z',\varphi'|z,\varphi)_i$$=$$(1/ 4 \pi N) A(z',\varphi'|z,\varphi)_i$.


To implement the rejection-free algorithm,
we start by computing $\lambda$ for this dynamic
\begin{equation}
  \lambda = {1 \over N} \sum_{i=1}^N \lambda_i \; ,
\end{equation}
\begin{equation}
{\small 
  \label{lambdaiH}
\begin{array}{lll}
  \lambda_i &:=& \int_{z'=-1}^1 \int_{\varphi'=0}^{2\pi} T(z',\varphi'|i)
  \left[ 1-A(z',\varphi'|z,\varphi)_i \right] d\varphi' dz' \nonumber \\
  &=& {(z+1) \over 2} - {k_BT \over 2S_i} \left[1-\exp(-(z+1)S_i/k_BT) \right] \; .
\end{array}
}
\end{equation}
Finally, the value of $t$ to exit from the current state is computed from Eq.~(\ref{t}).

According to Eq.~(\ref{C}),
\begin{equation}
{\small
  \label{CH}
  C(z',\varphi'|z,\varphi)_i = 
  {1 \over 4 \pi N (1-\lambda)} A(z',\varphi'|z,\varphi)_i \; ,
}
\end{equation}
but for a system with continuous degrees of freedom, 
it is not possible to group the movements into classes by energy changes, 
since these values form a continuous set,
and instead of grouping by energies,
we group the movements by {\it sites}. 
The probability to choose a site $i$ is
\begin{equation}
{\small 
  \label{Cix}
  C(i|x) = \int_{\rm sphere}
  C(z',\varphi'|z,\varphi)_i \; dz' d \varphi'
    ={1-\lambda_i \over N(1-\lambda)} \; .
}
\end{equation}
The search to find which spin $i$ changes can be performed with the
same integral-generator strategy described above.

Next, we choose $(z',\varphi')$ for the site $i$,
but since these are continuous variables, 
their distribution is described by a pdf that, 
according to the Bayes relation, 
is given by
\begin{equation}
  C(z',\varphi'|i)={1 \over 4\pi(1-\lambda_i)} A(z',\varphi'|z,\varphi)_i \; .
\end{equation}
Since this expression is independent of $\varphi'$, 
this coordinate is uniformly distributed on the interval $[0,2\pi)$. 
In contrast, $z'$ must be generated with pdf
\begin{equation}
{\small  
   \label{fz2}
    f^i_z(z')= \left\{ \begin{array}{ll}
        {1 \over 2(1-\lambda_i)} \exp((z'-z)S_i/k_BT) & \mbox{\rm ; if } z'<z \\
        {1 \over 2(1-\lambda_i)} & \mbox{\rm ; otherwise }
      \end{array} \right. \; .    
}
\end{equation}
A rejection-free random generator with this distribution can be constructed
by means of the same integral-generator strategy but on a continuous variable. 
First, define the integral
\begin{equation}
{\small
\begin{array}{lll}
    Q_i(z') &:=& \int_{z''=-1}^{z'} f^i_z(z'') dz''\\
    &=&  \left\{ \begin{array}{l}
        {k_BT \over 2(1-\lambda_i)S_i} 
        \left[ \exp((z'-z)S_i/k_BT) - 1 \right] + \Omega_i(z) \\
        \begin{array}{ll} 
          & \mbox{\rm ; if } z'\le z \\
          \Omega_i(z) + {(z'-z) \over 2(1-\lambda_i)} 
          & \mbox{\rm ; otherwise }
        \end{array}
      \end{array} \right. \; ,
\end{array}
}
\end{equation}
with $\Omega_i(z)$$=$$1-(1-z)/(2(1-\lambda_i))$.
Next, generate a random number $\tilde{r}$ uniformly distributed on $(0,1)$ and take $z'$ such that $Q_i(z')$$=$$\tilde{r}$,
\begin{equation}
{\small
    \label{GenZ}
    z'= \left\{ \begin{array}{l}
        z+ {k_BT \over S_i} 
        \ln{ \left[ 2(1-\lambda_i)S_i(\tilde{r}-\Omega_i(z))/k_BT+1 \right]
          } \\
        \begin{array}{ll} 
          & \mbox{\rm ; if } \tilde{r} \le \Omega_i(z)
          \\ z+2(1-\lambda_i)(\tilde{r}-\Omega_i(z))
          &\mbox{\rm ; otherwise}
        \end{array}
      \end{array} \right. \; .
}
\end{equation}
Finally, the new values $(z',\varphi')$ are transformed back into
Cartesian coordinates and rotated back to the original reference axis. 
Thus both the update time, $t$, and the exit state, $x'$, are chosen,
and the rejection-free implementation for a system with continuous degrees
of freedom is complete.
 
\label{Results}

We tested our rejection-free algorithm by computing the mean lifetimes
of a metastable state for an anisotropic Heisenberg model with $J_x$$=$$J_y$$=$$1.0$ and $J_z$$=$$2.0$ 
on a simple cubic (SC) lattice of size $10$$\times$$10\times$$10$ with periodic boundary conditions.  
The system is initially in a metastable state with all spins pointing in the $-z$ direction
with the external field in the opposite direction,
and all simulations were performed at temperatures $T < T_{\rm c}$,
where we have found the approximate value of the critical temperature to be $T_{\rm c}$$\simeq$$3.15J_x$.
Here, we concentrate on dynamic quantities,
such as the metastable escape time or lifetime, $\tau$, 
which is the number of Monte Carlo steps per site (MCSS) needed to obtain a magnetization $M_z$$=$$0$.  
Preliminary results from the algorithm for static quantities have been presented elsewhere \cite{Athens2000}.

To improve the efficiency of the site selection portion of the algorithm,
the sites were grouped into super-classes of lines and planes and organized into a three-level tree \cite{TreeSearch}.
Even with this improved tree-search, 
the steps of the rejection-free algorithm require more computational time
that the steps of a direct Metropolis implementation.
On average, the rejection-free method took $8.769$$\mu$s to perform one change,
and the direct Metropolis implementation took $1.162$$\mu$s to make one trial.
However, when many trials are required to make a single change,
the rejection-free algorithm is computationally more efficient than the direct implementation.

\begin{figure}
\includegraphics[width=\columnwidth]{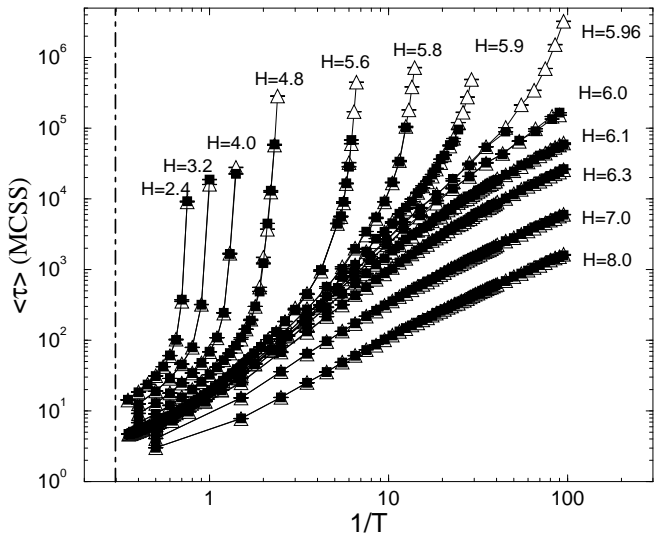}
\caption[]{
Average lifetimes, $\langle \tau \rangle$, of the metastable state of an anisotropic Heisenberg model 
on a $10^3$ cubic lattice at many magnetic field values.
Each point represents an average over 100 independent escapes.
Results for both the rejection-free algorithm (triangles)
and the direct implementation (filled squares) are shown.
The vertical line represents the approximate value of $T_{\rm c}$.

}
\label{FigTescapes}
\end{figure}

Figure \ref{FigTescapes} shows the average lifetimes, $\langle \tau \rangle$, of the metastable state 
computed by both rejection-free and direct implementations
vs.\ $1/T$ for many values of the external magnetic field.
As expected, the two methods give identical results for all temperatures and fields,
regardless of the switching mechanism 
(nucleation regime $H_z$$<$$6.0$, strong-field regime $H_z$$>$$6.0$).
See Ref.~\cite{NovotnyIJMPC} for a description of the switching mechanisms.

\begin{figure}
\includegraphics[width=\columnwidth]{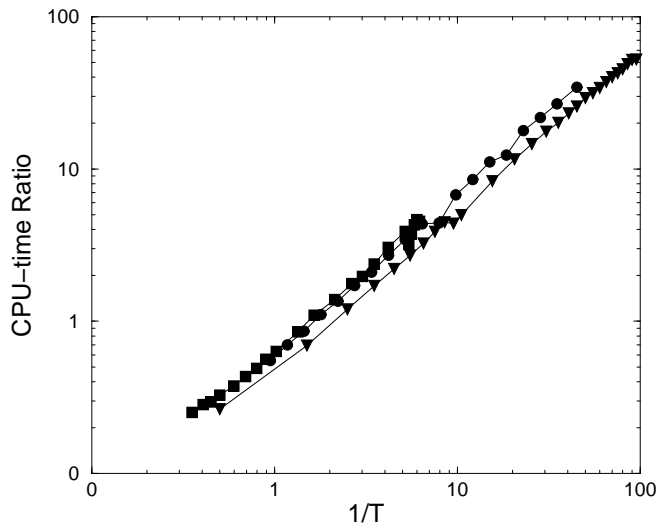}
\caption[]{
CPU time ratio vs.\ $1/T$.
The external field values are $H_z$$=$$5.6$ (squares), $H_z$$=$$5.96$ (circles), and $H_z$$=$$7.0$ (triangles).
}
\label{FigRatios}
\end{figure}

We define $\langle \tau \rangle_{\rm CPU, direct}$ as the average CPU time required to escape the metastable state,
when simulated by the direct implementation,
and $\langle \tau \rangle_{\rm CPU, rej-free}$ is similarly defined.
Figure \ref{FigRatios} shows the CPU-time ratio, $\langle \tau \rangle_{\rm CPU, direct}/\langle \tau \rangle_{\rm CPU, rej-free}$, vs.\ $1/T$.
For $T=1/100$, the rejection-free implementation is nearly two orders of magnitude faster than the direct implementation.

\label{Conclusions}

In summary, we have constructed a rejection-free Monte Carlo algorithm for a system with continuous degrees of freedom
which faithfully keeps the dynamic as compared to the direct implementation
but is orders of magnitude faster at low temperatures.
Our procedure is quite general and can be applied to other dynamics and other systems with continuous degrees of freedom.


The authors thank P.A. Rikvold for useful discussions. J.D. Mu\~noz
thanks H.J. Herrmann for hospitality. This work is partially supported
by the Deutscher Akademischer Austauschdienst (DAAD) through
scholarship A/96/0390 and the NSF through grant DMR-9871455.

\bibliography{apssamp}

\end{document}